\documentclass[letterpaper, 10 pt, conference]{ieeeconf}  % Comment this line out if you need a4paper

\IEEEoverridecommandlockouts                              % This command is only needed if 
\usepackage{graphicx}
\usepackage{mathptmx}
\usepackage{amsmath,amssymb,amsfonts}
\usepackage{algorithmic}
\usepackage{stfloats}
\usepackage{booktabs}% assumes new font selection scheme installed

\title{\LARGE \bf
An EEG Channel Selection Framework for Driver
Drowsiness Detection via Interpretability Guidance}

\author{Xinliang Zhou$^{1}$, Dan Lin$^{2}$, Ziyu Jia$^{3,4}$,  Jiaping Xiao$^{5}$, Chenyu Liu$^{1}$, Liming Zhai$^{2}$, and Yang Liu$^{1}$% <-this % stops a space
\thanks{*Corresponding Author: Liming Zhai.}% <-this % stops a space
\thanks{$^{1}$Xinliang Zhou, Chenyu Liu and Yang Liu are with the School of Computer Science and Engineering, Nanyang Technological University, 639798, Singapore {\tt\small\{xinliang001, chenyu003\}@e.ntu.edu.sg} {\tt\small yangliu@ntu.edu.sg}}%
\thanks{$^{2}$Dan Lin and Liming Zhai are with Continental-NTU Corporate Lab, Nanyang Technological University, 50 Nanyang Avenue, 639798, Singapore
        {\tt\small \ {\{dan.lin, liming.zhai\}@ntu.edu.sg}}}%
\thanks{$^{2,3}$Ziyu Jia is with the Brainnetome Center, Institute of Automation, Chinese Academy of Sciences, Beijing 100190, China, and also with University of Chinese Academy of Sciences, Beijing 100190, China
        {\tt\small \ {jia.ziyu}@outlook.com}}%
\thanks{$^{4}$Jiaping Xiao is with the School of Mechanical and Aerospace Engineering, Nanyang Technological University, 639798, Singapore
        {\tt\small \ {jiaping001}@e.ntu.edu.sg}}%
}

\def\BibTeX{{\rm B\kern-.05em{\sc i\kern-.025em b}\kern-.08em
    T\kern-.1667em\lower.7ex\hbox{E}\kern-.125emX}}

\begin{document}

\maketitle
\thispagestyle{empty}
\pagestyle{empty}

%%%%%%%%%%%%%%%%%%%%%%%%%%%%%%%%%%%%%%%%%%%%%%%%%%%%%%%%%%%%%%%%%%%%
\begin{abstract}

% Driver drowsiness has a crucial influence on driving safety, creating an urgent demand for driver drowsiness detection. Electroencephalogram (EEG) signal can accurately reflect the mental fatigue state and thus has been widely studied in driver drowsiness detection. However, the raw EEG data is inherently noisy and redundant, and existing works often use single-channel EEG data or full-head channel EEG data for model training, resulting in the limited performance of driver drowsiness detection. In this paper, we are the first to propose an Interpretability-guided Channel Selection (ICS) framework for the driver drowsiness detection task. Specifically, we design a two-stage training strategy to progressively select the key contributing channels with the guidance of interpretability. We first train a teacher network in the first stage using full-head channel EEG data. Then we apply the class activation mapping (CAM) to the trained teacher model to highlight the high-contributing EEG channels and further propose a channel voting scheme to select the top $N$ contributing EEG channels. Finally, we train a student network with the selected channels of EEG data in the second stage for driver drowsiness detection. Experiments are designed on a public dataset, and the results demonstrate that our method is highly applicable and can significantly improve the performance of cross-subject driver drowsiness detection.
Drowsy driving has a crucial influence on driving safety, creating an urgent demand for driver drowsiness detection. Electroencephalogram (EEG) signal can accurately reflect the mental fatigue state and thus has been widely studied in drowsiness monitoring. However, the raw EEG data is inherently noisy and redundant, which is neglected by existing works that just use single-channel EEG data or full-head channel EEG data for model training, resulting in limited performance of driver drowsiness detection. In this paper, we are the first to propose an Interpretability-guided Channel Selection (ICS) framework for the driver drowsiness detection task. Specifically, we design a two-stage training strategy to progressively select the key contributing channels with the guidance of interpretability. We first train a teacher network in the first stage using full-head channel EEG data. Then we apply the class activation mapping (CAM) to the trained teacher model to highlight the high-contributing EEG channels and further propose a channel voting scheme to select the top $N$ contributing EEG channels. Finally, we train a student network with the selected channels of EEG data in the second stage for driver drowsiness detection. Experiments are designed on a public dataset, and the results demonstrate that our method is highly applicable and can significantly improve the performance of cross-subject driver drowsiness detection.
\newline

\end{abstract}

\begin{figure*}[t] 
\centering 
%includegraphics[width=0.85\textwidth, height=0.35\columnwidth]{image/main_revise.pdf}
 \includegraphics[width=0.88\textwidth]{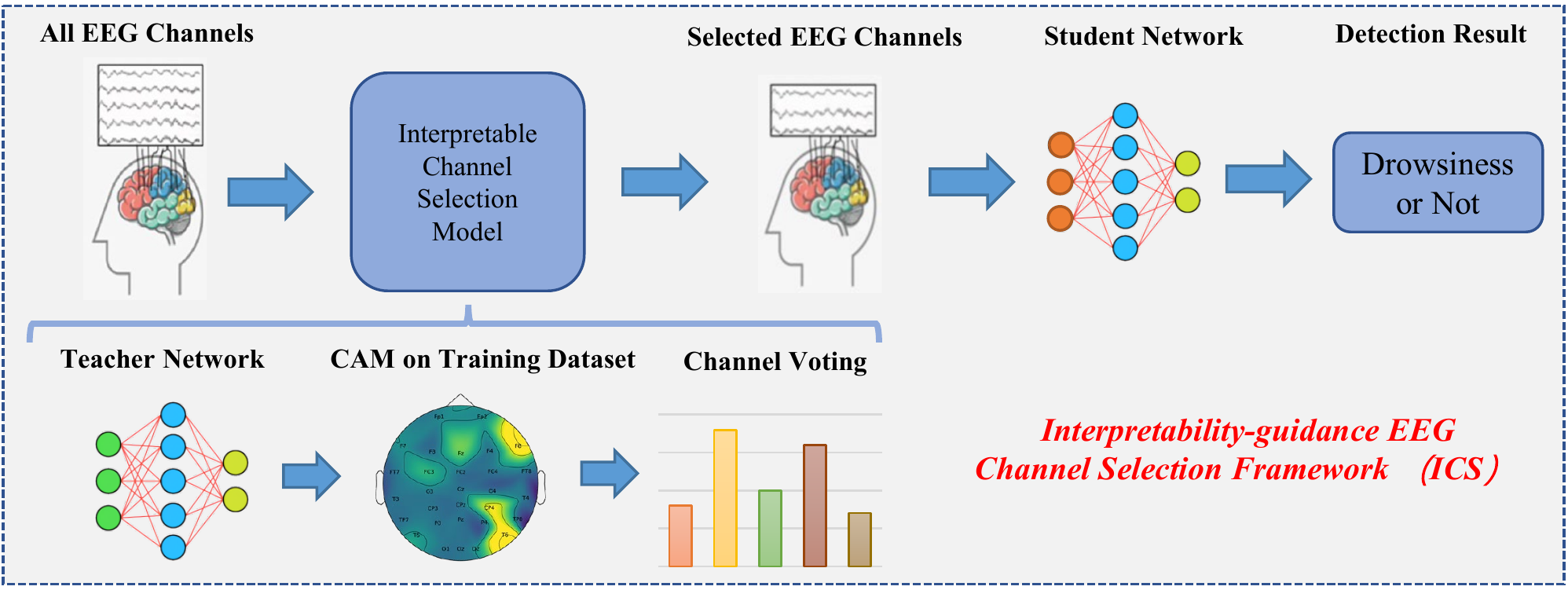}
\caption{The structure of the proposed ISC framework.} 
\label{structure}
\end{figure*}

%%%%%%%%%%%%%%%%%%%%%%%%%%%%%%%%%%%%%%%%%%%%%%%%%%%%%%%%%%%%%%%%%%%%%%%%%%%%%%%%
\section{INTRODUCTION}

Driver drowsiness has been a leading risk factor for traffic accidents and poses a major threat to roadway safety \cite{FatigueReport2021}. Therefore, developing high-performance drowsiness detection systems that continuously monitor the driver's state and alert the driver at the onset of drowsiness is quite crucial for safe driving.

Considerable efforts have been made to driver drowsiness detection based on various bio-traits, including eye blinking \cite{hsieh2013improved}, heart rate \cite{patel2011applying}, respiration \cite{tateno2018development}, and facial expression \cite{liu2020driver}. Among these bio-traits, the electroencephalogram (EEG), which directly measures brain activities, has gained more attention owing to its essential relationship with fatigue \cite{peng2022application}.
Current prevailing EEG-based driver drowsiness detection methods mainly utilize deep neural networks (DNNs) to learn feature representations directly from raw EEG data for binary classification (drowsiness or not).
The raw EEG data is obtained by multiple electrodes positioned on specific areas of the scalp and continuously capture the brain's electrical activity. Thus the EEG data is high-dimensional and usually contains noise, artifacts, and task-irrelevant information. The existing DNNs for drowsiness detection either use single-channel EEG data \cite{cui2022compact} or full-head channel EEG data \cite{cui2022eeg} as input data, resulting in limited performance due to insufficient learning (from single-channel with incomplete information) or biased learning (from the full-head channel with noises). Since data is the basis of training DNN models, it is highly demanded to study how the different channels of EEG data affect the DNNs and how to select suitable EEG channels for effective feature learning.

% To circumvent this dilemma, we present
% This paper presents an EEG channel selection framework for driver drowsiness detection from an interpretability point of view.
Although some EEG channel selection methods exist, they are all task-specific (designed for a particular task and cannot be successfully applied to drowsiness detection) and subject-specific (the selected channels vary with each subject and may not be suitable for other subjects).
To make the selected channels more general to the driver drowsiness detection task, we propose a two-stage training strategy that contains a teacher network and a student network to refine the feature learning with channel selection in a coarse-to-fine fashion. For the teacher network, we first use an interpretability method to analyze each channel's contribution to the drowsiness detection with training data and then design a voting scheme to select the top $N$ contributing channels according to the performance of the teacher network. The selected channels of EEG data are finally fed to the student network for further training. Experiments conducted on a public dataset demonstrate that our method is highly applicable to driver drowsiness detection and can significantly improve the detection accuracy on cross-subjects.

\section{Related Work}

\noindent {\bf {EEG-based Drowsiness Detection.}} EEG signals directly measure brain activity and reflect the fatigue state more straightforwardly than other types of signals (e.g., eye blinking, driver expression, and heart rate). In the pre-deep learning era, EEG-based drowsiness detection tasks focused more on extracting effective features. Chai et al. \cite{chai2016driver} utilized independent components by entropy rate bound minimization analysis (ERBM-ICA) for the source separation and autoregressive (AR) modeling for extracting the features. Gao et al. \cite{gao2019eeg} proposed a multileveled feature generator that combines a one-dimensional binary pattern and statistical features. Schwendeman et al. \cite{schwendeman2022drowsiness} applied Welch’s method to extract the power spectral density (PSD) as drowsiness features from the dry electrode in-ear EEG. 
Recently, deep learning has been adopted prevalently for EEG-based drowsiness detection. Nissimagoudar et al. \cite{nissimagoudar2019deep} and Ding et al. \cite{ding2019cascaded} applied DNNs to single-channel EEG data to detect driver drowsiness. Furthermore, Cui et al. \cite{cui2022eeg} proposed an InterpretableCNN with two convolutional layers, one point-wise convolution layer, and one depth-wise convolution layer to process multi-channel EEG data and construct a more robust detection system. Different from these methods that rely on insufficient or noisy data, we select more informative data from raw data for model training.

\noindent {\bf {Channel Selection.}} 
In previous studies, some scholars formulated the EEG channel selection into mathematical and statistical problems. With the help of the non-dominated sorting genetic algorithm (NSGA), Luis Alfredo Moctezuma et al. presented a channel optimation method for epileptic-seizure classification. Alyasseri et al. \cite{alyasseri2020person} proposed hybrid optimization techniques based on the binary flower pollination algorithm (FPA) and $\beta$-hill climbing (called FPA $\beta$-hc). Varsehi et al. \cite{varsehi2021eeg} assumpted that there are causal interactions between the channels of EEG signals, and applied Granger Causality (GC) analysis to verify this assumption. Besides, reducing the number of channels can effectively minimize the computational complexity and training time. To this end, Yang et al. \cite{yang2021grouped} proposed a grouped dynamic EEG channel selection method, GDCSBR, which provided flexibility for subsequent dynamic channel selection. Differently, Zhang et al. \cite{zhang2021motor} utilized a sparse squeeze-and-excitation module, and Gaur et al. \cite{gaur2021automatic} proposed using the Pearson Correlation Coefficient (PCC), achieving automatic subject-specific selection, respectively.
These methods are based on learning schemes or statistical schemes but are unsuitable for driver drowsiness detection. In contrast, we use interpretability to guide the channel selection for drowsiness detection.

\section{Methodology}

\subsection{Framework overview}
\label{sec:format}
Our interpretability-guided channel selection (ICS) framework for EEG data is shown in Fig. \ref{structure}. To select useful EEG channels, the ICS uses a two-stage training strategy. In the first training stage, the teacher network takes as input all EEG channels and is optimized for the drowsiness detection task. Then the class activation mapping (CAM) \cite{zhou2016learning} is applied to each training sample to show the importance of different EEG channels, following which the channel voting is performed to select the contributing channels. In the second stage, the student network is trained with the selected EEG channels and is used as the final detection model.
The teacher network and the student network have similar architectures, except that the convolution channels in the first layer of the student network are adjusted with the number of selected EEG channels.

\subsection{Interpretability Guidance}

Besides the two-stage networks, the interpretation technique, CAM  method, plays a crucial role in the ICS framework. The CAM method is an effective interpretation technique that can pinpoint the discriminative regions of each input sample for a trained Convolutional Neural Networks (CNN) model. In particular, a heatmap is generated from the activations following the final convolutional layer for each input sample. The map is then interpolated to the size of the input sample to reveal what extent the input sample's local regions contribute to the classification.

However, the CAM method cannot be used directly for time-series EEG data and our model, because it was initially only intended for 2-D image data and deep CNN structure with a global average pooling (GAP) layer. Inspired by Cui et al.'s work \cite{cui2022eeg}, we modify the original CNN by changing the final dense layer to the GAP layer and applying the CAM method to visualize the EEG classification. The visualized heatmap can verify the contribution of each channel to the classification. After modification, the two-stage network architecture integrated with the CAM technique can outperform the original detection system's performance.
% \begin{figure}[t] 
% \centering 
% %\includegraphics[height=0.73\columnwidth,width=0.83\columnwidth]{xxxx.pdf} 
% %\includegraphics[width=0.83\columnwidth]{image/voting.pdf} 
% \includegraphics[width=1.0\columnwidth]{image/xxxx.pdf} 
% \caption{Visualization of CAM and voting scheme. The normalized heatmap score represents the level of contribution. The higher the score (indicated by the lighter color), the more significant the contribution. The voting can be divided into two processes, sum up and sort. Sum and sort high-contribution samples' heatmap scores, then we can obtain the channel contributions.}
% \label{voting}
% \end{figure}

\subsection{Voting Scheme}

In this section, we detail how the voting scheme works. To select the top $N$ channels that contribute significantly to the system's classification result, we apply CAM on the training set to calculate the contribution of each channel to the classification. We pick up the input sample points with high confidence levels. Specifically, the probability of the model's prediction for both categories (drowsiness or not) is obtained by the softmax layer

\begin{equation}
\sigma(\mathbf{z})_i=\frac{e^{z_i}}{\sum_{j=1}^K e^{z_j}} \quad, \text { for } i=1, \ldots, K,
\end{equation}
where $\mathbf{z}$ is the intermediate tensor inside the model before feeding into the softmax layer, $e^{z_i}$  donates the standard exponential function for the input tensor, and $K$ is the number of prediction classes in the teacher network ($K$ = $2$ in our task). Hence, we can obtain the prediction probabilities of two classes.

\begin{figure}[t] 
\centering 
\includegraphics[width=0.9\columnwidth]{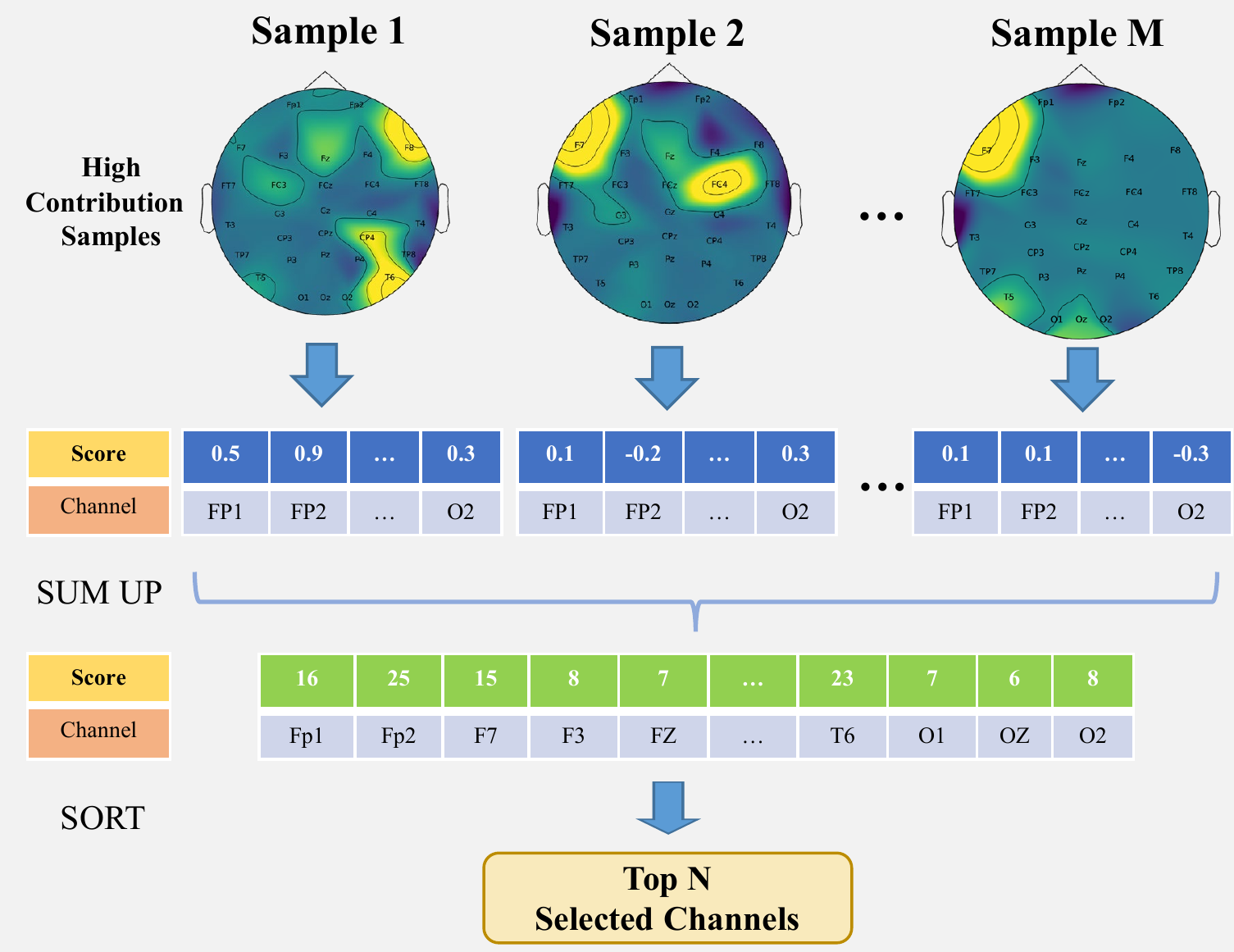} 
\caption{Visualization of CAM and voting scheme. The normalized heatmap score represents the level of contribution. The higher the score (indicated by the lighter color), the more significant the contribution. The voting can be divided into two processes, sum up and sort. Sum and sort high-contribution samples' heatmap scores, then we can obtain the channel contributions.}
\label{voting}
\end{figure}

We consider the sample points in channels as high-contribution for the model classification when and only when the model predicts correctly and the prediction softmax probability is greater than or equal to $0.90$.
We conduct a prediction with the teacher model for each sample point and then apply the CAM method to the entire training dataset to visualize the EEG classification during the training process. The visualized heatmap, as shown in Fig. \ref{voting}, demonstrates the contribution of each channel. We select the high-confidence sample points in the next step based on the aforementioned selection rule.

After we obtain the high-contribution sample points, we need to set them to make voting trace back to the top $N$ contribution channels. To use the global information, we sum up the heatmap scores of each channel for each sample and then rank all the channels in descending order. As such, we obtain the ranking of all channels in terms of their contribution to the correct model classification. Next, we select the top $N$ channels according to the ranking. With high-confidence samples on the training dataset and voting scheme, our approach can select the top $N$ channels that are most closely associated. Finally, the student network are retrained using the the top $N$ channels and thus can improve the model performance.

\section{Experiments}

\label{sec:pagestyle}

\subsection{Experimental Setup}
\subsubsection{Dataset}
% A public EEG dataset  was used in the study. The dataset was collected from 27 subjects (aged from 22 to 28) in a sustained-driving task in a virtual reality simulator. In the process, lane-departure events were randomly introduced which drifted the car away from the central lane. The participants were required to respond immediately to the events by driving the car back to central lane. The drowsiness level can be reflected by how fast the subjects respond to the events.

% In this paper, a public EEG dataset \cite{cao2019multi} was utilized. The data was collected from 27 subjects (aged from 22 to 28) who performed a sustained driving task in a virtual reality simulator. The participants were required to immediately respond to the random lane-departure events by returning the vehicle to the center lane. The label of alert and drowsiness state can be determined by the subjects' reaction times. 
% We followed the previous work   \cite{cui2022eeg}\cite{wei2018toward} and got a mini version of the dataset, which contains 11 subjects and 2952 samples,  and we call it a mini driver drowsiness dataset (MDDD). Each sample $X \in R^{30 \times 384}$, where 30 stands for the number of channels and 384 stands for the number of sample points.
We perform the drowsiness detection experiments on a public EEG dataset \cite{cao2019multi}. The data was collected from 27 subjects who completed a lengthy driving task in a virtual reality simulator. The participants were required to immediately return the vehicle to the center lane in response to random lane departure events. The label (drowsiness or not) can be determined based on the subjects' reaction times.
We follow the previous works \cite{cui2022eeg,wei2018toward} and build a mini version of the dataset containing 11 subjects and 2952 samples, and we call it a mini driver drowsiness dataset (MDDD). Each sample, $X \in R^{30 \times 384}$, where 30 and 384 are the number of channels and sample points, respectively.

\subsubsection{Baselines}
To demonstrate the superiority of the proposed ICS framework, we compare with four typical CNN-based methods in EEG classifications, including EEGNet \cite{lawhern2018eegnet} (both EEGNet 4,2 and EEGNet 8,2), ShallowConvNet \cite{schirrmeister2017deep} and InterpretableCNN \cite{cui2022eeg}. All the baselines are implemented using their default settings.

\subsubsection{Training and Testing}
We conduct the training and testing with the Leave-one-subject-out analysis. In particular, the classifiers are trained using the EEG data from all other subjects and tested using the data from just one subject. Every subject is used as a test subject once during each iteration. 
The performance of drowsiness detection is measured by detection accuracy, which is the ratio of the number of correctly predicted testing samples to the number of all testing samples.

\begin{table}[t]
\caption{ Mean accuracies on MDDD with the changes in the numbers of top channels selected.}
\centering
\scalebox{0.84}{
\begin{tabular}{@{}ccccccc@{}}
\toprule
Nums of Channels & $N$ = $5$  & $N$ = $10$          & $N$ = $15$ & $N$ = $20$ & $N$ = $25$ & $N$ = $30$ \\ \midrule
EEGNet 4,2 \cite{lawhern2018eegnet}       & 0.6742 & \textbf{0.7296} & 0.7283 & 0.7121 & 0.7022 & 0.6831 \\
EEGNet 8,2 \cite{lawhern2018eegnet}       & 0.6798 & \textbf{0.7369} & 0.7342 & 0.7155 & 0.6977 & 0.6842 \\
ShallowConvNet \cite{schirrmeister2017deep}   & 0.7741 & \textbf{0.7944} & 0.7881 & 0.7821 & 0.7725 & 0.7665 \\
InterpretableCNN \cite{cui2022eeg} & 0.7785 & \textbf{0.8138} & 0.8051 & 0.8012 & 0.789  & 0.7826 \\ \bottomrule
\end{tabular}}
\label{compare}
\end{table}

\begin{table*}[t]
\centering
\caption{The comparison of the accuracies (Standard Deviation) of MDDD between the original four classifiers and them after applying ICS for selecting the top 10 channels.}
\vspace{-5pt}
\resizebox{\textwidth}{!}
{
\scalebox{1.0}{
\begin{tabular}{@{}ccccccccc@{}}

\toprule
Method     & \multicolumn{2}{c}{EEGNet 4,2 \cite{lawhern2018eegnet}} & \multicolumn{2}{c}{EEGNet 8,2 \cite{lawhern2018eegnet}} & \multicolumn{2}{c}{ShallowConvNet \cite{schirrmeister2017deep}} & \multicolumn{2}{c}{InterpretableCNN \cite{cui2022eeg}} \\ \midrule
subject ID & Original Acc & ICS Acc       & Original Acc & ICS Acc       & Original Acc   & ICS Acc         & Original Acc    & ICS Acc          \\ \cmidrule(l){2-9} 
0          & 0.8352       & \textbf{0.8455} & 0.8251       & \textbf{0.8312} & 0.8228         & \textbf{0.8422}   & 0.8457          & \textbf{0.8617}    \\
1          & 0.3125       & \textbf{0.5333} & 0.4525       & \textbf{0.6033} & 0.8018         & \textbf{0.8226}   & 0.7121          & \textbf{0.8939}    \\
2          & 0.5675       & \textbf{0.6021} & 0.5528       & \textbf{0.7021} & 0.6446         & \textbf{0.7612}   & \textbf{0.8333}          & 0.8133             \\
3          & 0.8521       & \textbf{0.8666} & 0.8024       & \textbf{0.8228} & 0.7026         & \textbf{0.7643}   & 0.7770           & \textbf{0.8041}             \\
4          & 0.4571       & \textbf{0.6051} & 0.4563       & \textbf{0.6017} & \textbf{0.8308}         & 0.8128            & 0.8661          & \textbf{0.8929}    \\
5          & 0.4776       & \textbf{0.6188} & \textbf{0.5076}       & 0.4522          & 0.7912         & \textbf{0.8225}   & \textbf{0.8916}          & 0.8313             \\
6          & \textbf{0.8663}       & 0.8443          & \textbf{0.8218}       & 0.8011          & 0.7313         & \textbf{0.7757}   & 0.6176          & \textbf{0.6373}    \\
7          & \textbf{0.7665}       & 0.7227          & 0.7435       & \textbf{0.7882} & \textbf{0.7122}         & 0.6883            & \textbf{0.7765}          & 0.6970              \\
8          & 0.6998       & \textbf{0.7127} & 0.7323       & \textbf{0.7943} & \textbf{0.8988}         & 0.8873            & \textbf{0.8694}          & 0.8631             \\
9          & 0.8781       & \textbf{0.8521} & 0.8328       & \textbf{0.8844} & 0.8308         & \textbf{0.8067}   & 0.6759          & \textbf{0.8519}    \\
10         & 0.8011       & \textbf{0.8225} & 0.7994       & \textbf{0.8476} & 0.6643         & \textbf{0.7548}   & 0.7434          & \textbf{0.8053}    \\ \midrule
Mean       & 0.6831 (0.1971)       & \textbf{0.7296} (0.1232) & 0.6842 (0.1575)      & \textbf{0.7390} (0.1324) & 0.7665 (0.080)        & \textbf{0.7944} (0.053)  & 0.7826 (0.027)         & \textbf{0.8138} (0.008)   \\ \bottomrule
\label{result}
\end{tabular}}}
% \caption{The comparison of the accuracies (Standard Deviation) of MDDD between the original four classifiers and them after applying ICS for selecting the top 10 channels.}
\end{table*}

\subsection{Ablation Study}
The values of $N$ in the ICS framework are adjusted to select the most suitable $N$ value for the performance. The results are shown in Table \ref{compare}. We can see that the best performance of the four classifiers is achieved with $N$ = $10$. When the number of selected channels is too small, the classifiers can not perform very well because they do not have enough information. As the value of $N$ increases (from $10$ to $30$, with  a step size of $5$), the classifier's performance decreases since there are too many noises in the newly added channels, and the classifier incorrectly learns the features, hence decreasing the model performance.

\subsection{Comparison with Previous Methods}

We apply our ICS framework to existing driver drowsiness detection methods and show the results in Table \ref{result}. Note that we use $N = 10$ for channel selection for the following experiments. We can see from Table \ref{result} that the proposed ICS framework provides a significant contribution to improving the previous models' performance in driver drowsiness detection. Specifically, the original EEGNet 4,2 and EEGNet 8,2 have a similar performance in terms of average accuracy, reaching around $68.3\%$ and $68.4\%$. After making the channel selection with ICS, their average accuracies improved significantly, from $68.31\%$ to $72.96\%$ and $68.42\%$ to $73.69\%$, respectively. As for ShallowConvNet and InterpretableCNN, their average accuracies increase $2.79\%$ and $3.12\%$, respectively.
In addition to improving the average accuracy, ICS also greatly boosts the accuracy of each subject. In terms of the Leave-one-subject-out cross-validation on the four classifiers, $70.45\%$ (31 out of 44) of the subjects' detection accuracy has been improved). 

Apart from the accuracy improvement, the ICS framework can also effectively improve the stability of the classifier. Specifically, for the two EEGNet classifiers, the application of ICS decreases their standard deviation of detection accuracy for each test subject from $0.1971$ to $0.1232$ and from $0.1575$ to $0.1324$, respectively. In contrast, the original ShallowConvNet and InterpretableCNN already have higher stability in detection accuracy, with standard deviations of $0.080$ and $0.088$, respectively.
The improvement of their stability via ICS is limited, with a decrease of $0.027$ and $0.008$, respectively. Hence, we can conclude that ICS can help the model to learn more informative features and thus improve the robustness of driver drowsiness detection.

\subsection{Comparison with other Channel Selection Schemes}
\begin{figure}[t] 
\centering 
\includegraphics[width=1.0\columnwidth]{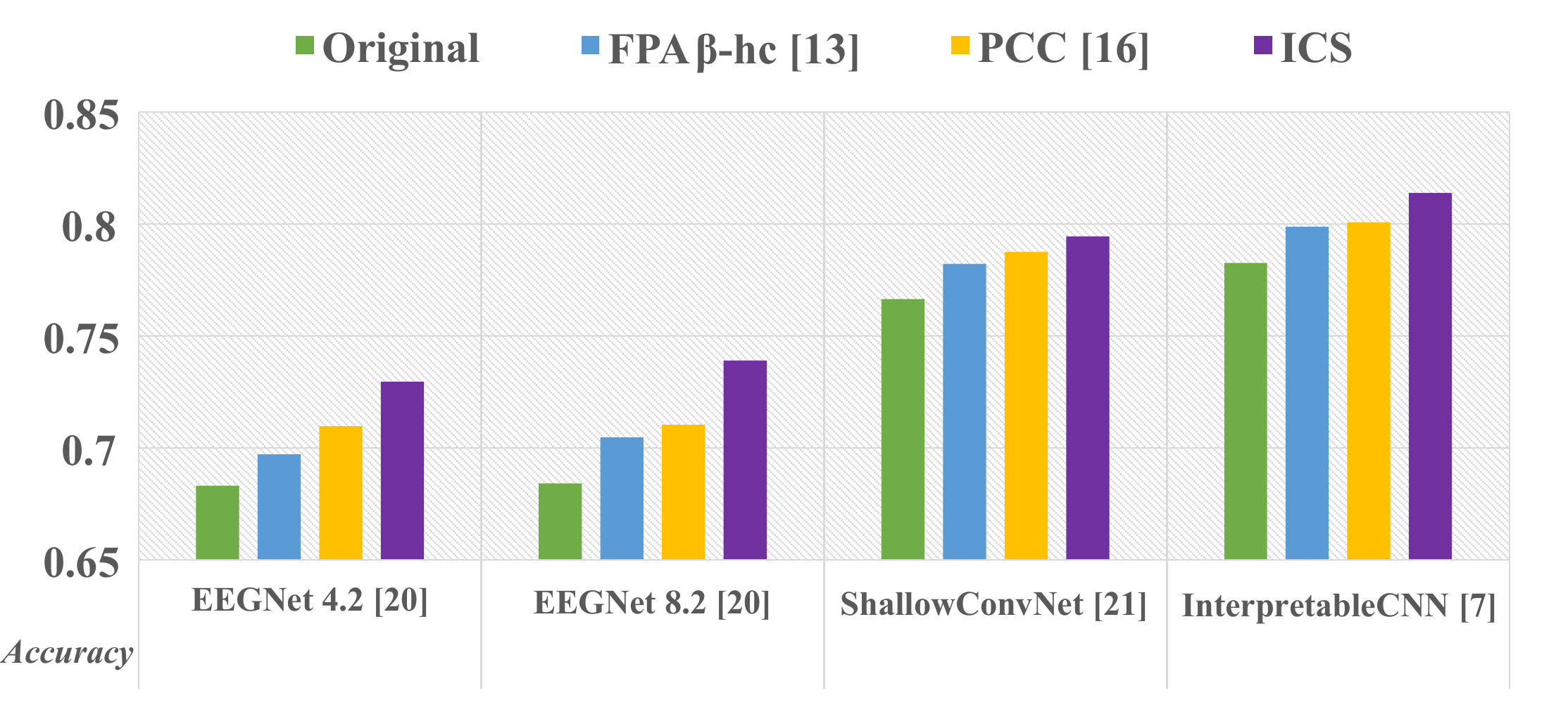}
\caption{Comparison of ICS and other typical channel selection methods.} 
\label{sota} %用于文内引用的标签
\end{figure}

% In this section, we visualize the improvement of driver drowsiness detection with the help of ICS and two other SOTA channel selection methods. The result demonstrates that the previous SOTA method, PCC and ACS, can improve the system's performance by around $1.5\%$-$2.0\%$. In contrast, ICS outperforms the previous method, which provides an average improvement at around $4.0\%$.
In this section, we compare our ICS with other typical channel selection methods to see how channel selection contributes to the drowsiness detection task. The results are shown in Fig. \ref{sota}, which demonstrates that previous typical methods, \emph{i.e.}, FPA $\beta$-hc, and PCC, can only slightly improve the detection accuracy by $1.5\%$-$2.0\%$. In contrast, the ICS significantly outperforms the previous methods, achieving an average improvement of around $4.0\%$.

% \subsection{Advantages and Limitations}
% Compared to previous EEG channel selection methods, the channels selected by ICS are more intuitive and consistent with what people have previously said they know about neuroscience \cite{kong2017assessment}. As for the limitations, the ICS is not an end-to-end framework, which means that it requires two training sessions for each implementation, which is time-consuming.

% To be mention that, although the dataset has been well preprocessed, state-of-the-art models are very prone to be overfitting. 

% the Traditional methods like SVM perform poorly due to the low signal-to-noise ratio and the variability of EEG. Deep learning models, such as ShallowNet and MCNN, outperform the traditional method by achieving 72.9\% and 75.1\% accuracy, respectively. However, these two methods can’t overcome the differences across domains. The ShallowNetTL uses some data from the target domain for fine-tuning to achieve transfer of the model from the source domain to the target domain with a classification accuracy of 76.8\%. DARA uses a domain adaptive approach to reduce the domain variance and achieve higher performance. Ablation experiments on MSTAN demonstrate the effectiveness of extracting multilevel spatial-temporal features and narrowing the difference between domains. The ability of the proposed MSTAN to extract domain-invariant multi-level spatial-temporal features allows it to achieve a higher accuracy of 79.2\%.

\section{CONCLUSIONS}

This paper proposes an effective channel selection framework for driver drowsiness detection with the guidance of interpretability. Based on the interpretable CAM algorithm and the voting scheme, the proposed ICS can effectively select the most relevant channels and remove the channels that do not contribute positively to drowsiness detection. The ICS is an interpretability-guidance approach to cross-subject channel selection, taking advantage of interpretability and selecting the channels intuitively and transparently. The experiment on a public dataset demonstrates that our framework is universally applicable and can significantly improve the performance of previous models.

As for the limitations, the ICS is not an end-to-end framework, which means that it requires two training stages and is thus time-consuming. In the future, we intend to study the efficient optimization of teacher network and student network in a unified manner. We will further apply our ICS to other EEG scenarios, such as EEG epilepsy detection.

\section{ACKNOWLEDGMENTS}
This research is partially supported under the A). RIE2020 Industry Alignment Fund – Industry Collaboration Projects (IAF-ICP) Funding Initiative, as well as cash and in-kind contribution from the industry partner(s). B). National Research Foun-dation NRF Investigatorship NRF-NRFI06-2020-0001, and C). the National Research Foundation Singapore and  DSO National Laboratories under the AI Singapore Programme (AISG Award No: AISG2-RP-2020-019).

% \end{thebibliography}
\bibliographystyle{IEEEbib}
\bibliography{ref}

\end{document}